\documentclass[atmp]{ipart_v1}

\Vol{19}
\Issue{3}
\Year{2015}
\firstpage{701}

\usepackage{t1enc}
\usepackage[latin1]{inputenc}
\usepackage[english]{babel}

\usepackage{amsthm}
\usepackage{yfonts}

\usepackage{bbm}
\usepackage{bm}
\usepackage{mathrsfs}
\usepackage[enableskew]{youngtab}

\newcommand{\be}[0]{\begin{equation}}
\newcommand{\ee}[0]{\end{equation}}

\numberwithin{equation}{section}

\theoremstyle{plain}

\begin{document}

\title[The Schur-Weyl duality in the one-dimensional Hubbard model.]{The application of the Schur-Weyl duality in the one-dimensional Hubbard model.}

\author[Dorota Jakubczyk]{Dorota Jakubczyk}

\begin{abstract}
We present the application of the Schur-Weyl duality in the one-dimensional Hubbard model in the case of half-filled system of any number of atoms. We replace the actions of the dual symmetric and unitary groups in the whole $4^{N}$- dimensional Hilbert space by the actions of the dual groups in the spin and pseudo-spin spaces. The calculations significantly reduce the dimension of the eigenproblem of the one-dimensional Hubbard model.
\end{abstract}

\maketitle

\section{Introduction and summary}

One of the most successful descriptions of electrons in solids is band theory. It is based on reducing many-body interactions to an effective one-body description, i.e., on neglecting the two-body potential. The Hubbard model \cite{Hubbard} became especially important as it showed that for half-filling the Mott transition is reproduced, that could not be understood in terms of conventional band theory. John Hubbard (1931-1980) found the model to be the simplest that produces both a metallic and an insulating state of approximate behaviour of interacting electrons in a solid, depending on the value of on-site repulsion $u$. It has been used for more than a half a century in attempts to describe the electronic properties of solids with narrow bands, band magnetism (iron, cobalt, nickel), the Mott metal-insulator transition, electronic properties of high-Tc cuprates \cite{essler, Gebhard}. The Hubbard model is an extension of the so called tight-binding model, where electrons can hop between lattice sites as independent particles. 

The model can be explored in obviously two limiting cases, i.e. strong coupling approximation when hopping is suppressed $t<<u$ and when there are no interactions $u<<t$. The first case provides the half-filled system to be an insulator, whereas the latter case turn out to be the metalic since the Hamiltonian can be solved in $k$-space.

The aim of the calculations presented in the present paper is to determine the eigenbasis of the spin and the pseudo-spin symmetries using the Schur-Weyl duality (SWD)\cite{jak2018, qudit} resulting in - for the half-filling case - a significance simplifications of the eigenproblem of the one-dimensional Hubbard Hamiltonian. SWD was introduced by Schur \cite{Schur} and then further developed by Weyl \cite{Weyl}, who showed that Young symmetrizators of symmetric group can be used to obtain irreducible representations of a unitary group. In order to calculate the irreducible basis of these dual groups for any number of atoms in the spin or pseudo-spin space we use the technique of Jucys-Murphy operators presented for example in previous papers of the author - for example in \cite{per}.

\section{The model}

The dynamics of the finite set of interacting electrons, occupying the one-dimensional chain, consisted of $N$ atoms, can be described by the Hubbard Hamiltonian in the following form
\be\label{11}
{H}=t\sum_{i\in \tilde{2}}\sum_{j\in \tilde{N}}({c}^{\dagger}_{ji}{c}_{j+1 i}+{c}^{\dagger}_{j+1 i}{c}_{ji})+u\sum_{j\in \tilde{N}}{n}_{j \,\,+}{n}_{j\,\,-},
\ee
where $\tilde{N}=\{j=1,2,\dots , N\}$ denotes the set of atoms of the chain, $\tilde{2}=\{i=+,-\}$ denotes the set of spin projections, ${n}_{ji}={c}^{\dagger}_{ji}{c}_{ji}$, and finally ${c}^{\dagger}_{ji}$ , ${c}_{ji}$ are the canonical Fermi operators, that is creation and anihilation operators of electron of spin $i$, on the site $j$ with conventional fermionic anticomutational relation, namely
\be\label{12}
c_{k\alpha}c_{m\beta}^{\dagger}+c_{m\beta}^{\dagger}c_{k \alpha}=\delta_{km}\delta_{\alpha \beta},
\ee
\be\label{13}
c_{k\alpha}c_{m\beta}+c_{m\beta}c_{k \alpha}=0.
\ee
The electron hopping in the Hubbard Hamiltonian can only take place between nearest-neighbour sites, and all hopping processes have the same kinetic energy.\\
The set of all linearly independent vectors called \emph{electron configurations} \cite{jakubczyk1} provides the initial, orthonormal basis of the Hilbert space $\mathcal{H}$. These configurations are defined by the following mapping
\be\label{12}
f:\tilde{N}\longrightarrow \tilde{4},
\ee
and constitute the $N$-sequences of the elements from the set $\tilde{4}=\{\pm, \emptyset , +, -\}$
\be\label{13}
|f>=|f(1)f(2)\dots f(N)>=|i_{1} i_{2} \dots  i_{N}>,\,\, i_{j}\in \tilde{4},\,\, j\in \tilde{N},
\ee
where $\emptyset$ denotes the empty node, $+$ and $-$ stand for one-node spin projection equal to $\frac{1}{2}$ and $-\frac{1}{2}$, respectively, $\pm$ denotes the double occupation of the one node by two electrons with different spin projections,
with 
\be\label{14}
\mathcal{H} = lc_{\mathbb{C}} \,\, \tilde 4^{\tilde N},\,\,\,\tilde{4}^{\tilde N} = \{ f: \tilde N \longrightarrow \tilde 4 \}.
\ee
The number $\tilde 4^{\tilde N}$ follows from the fact that the four states
\be\label{15}
|\emptyset>,\,{c}^{\dagger}_{j+}|\emptyset>=|+>,\,{c}^{\dagger}_{j-}|\emptyset>=|->,\,{c}^{\dagger}_{j+}{c}^{\dagger}_{j-}|\emptyset>=|\pm>,
\ee
are associated with every lattice site $j\in \tilde{N}$.

\subsection{Symmetries of the model}

The symmetries of the one-dimensional Hubbard model has been systematically studied by many researchers, starting from Lieb and Wu \cite{Lieb}, Yang \cite{Yang1, Yang2} and continued in for example \cite{Yang3,Yang4}, with the book of Essler et al. \cite{essler} as the eminent sumation and suplement of their work. Since the periodic boundary condition are assumed, the Hamiltonian $(\ref{11})$ has the obvious translational symmetry $({c}_{N+1i}={c}_{1i})$, this mean that one-particle Hamiltonian of the form $(\ref{11})$ is completely diagonalised by a Fourier transformation in the form
\be\label{21}
c_{k}^{\dag}=\frac{1}{\sqrt{N}}\sum_{j\in \tilde{N}}\mbox{exp}(i2\pi kj/N)c_{j}^{\dag},\,\,k\in B,
\ee
where 
\be\label{22}
B=\{k=0, \pm1, \pm2,  \dots ,
\left
\{
\begin{array}{ll}
\pm(N/2-1), N/2, & \mbox{for N even} \\
\pm(N-1)/2, & \mbox{for N odd}
\end{array}
\right\}
\ee
and labels irreducible representations (irreps) \cite{jakubczyk1,homotopia}  
\be
\Gamma _{k}(j)=\sum_{j\in \tilde{N}}\mbox{exp}(i2\pi kj/N),\,\,j\in \tilde{N}.
\ee
of the translational symmetry group $C_{N}$.\\
In case of any number of electrons the \emph{$\kappa$-tuply rarefied orbits} $\mathcal{O}_{f^{i}}$ of the group $C_{N}$ can appear, where $f^{i}$ denotes the \emph{initial} electron configuration, and the positional index j can be restricted due to periodicity to the subset
\be
\tilde{N_{\kappa}}=\{j=1,2,\dots , N/\kappa\}\subset \tilde{N}.
\ee 
In order to obtain all the elements i.e. electron configurations of the orbit $\mathcal{O}_{f^{i}}$ of the translational symmetry group $C_{N}$ for any number of electrons, the action of the element $(C_{N})^{1}$ of the group $C_{N}$ is defined as follows
\be
(C_{N})^{1}c^{\dagger}_{ji}=c^{\dagger}_{(j+1)_{mod\,N}i}.
\ee
The Hamiltonian for any number of atoms $N$ can be introduced as the sum of the Hamiltonians depending on the values of the quasi-momenta k: 
\be\label{Hk}
{H}=\bigoplus_{k\in B}{{H}(k)},
\ee
and hence the Hilbert space $\mathcal{H}$ decomposes into subspaces as follows
\be
\mathcal{H}=\bigoplus_{k\in B}{\mathcal{H}^{k}},\,\,\,{\mathcal{H}^{k}}=lc_{\mathbb{C}} \,\,b_{w},
\ee
where $b_{w}$ is called the \emph{basis of wavelets} \cite{bazaprofili}.\\
The dimension of the subspace ${\mathcal{H}^{k}}$ can be less than $|B|$, since $\kappa$-tuply rarefied orbit $\mathcal{O}_{f^{i}}$ contributes only to the Fourier transformation of the quasi-momentum $k$ within the rarefied Brillouin zone $B/\kappa$
\be
B/\kappa =\{k\in B|k/\kappa \,\,\mbox{is an integer}\}\subset B.
\ee\\

Apart from the cyclic symmetry system reveals for the half-filling of the electrons, two independent $SU(2)$ symmetries \cite{essler, Cuoco}, that is $SU(2)\times SU(2)$ in the spin and pseudo-spin space, respectively. This symmetry involves spin and charge degrees of freedom, and are related with four elementary excitation, that is spinons with respect to the spin, and holon and antiholon, with respect to the charge. The set $\tilde{4}=\{+, -, \pm, \emptyset\}$ can be decomposed into two subsets, where first $\tilde{2'}=\{+,-\}$ reflecting the invariance of ${H}$ under the spin rotation is related with the left factor of the direct product $SU(2)\times SU(2)$ of the unitary groups of the system and the second set $\tilde{2''}=\{\pm, \emptyset \}$ is related with the right factor. Thus, one has two sets of generators $\{{S}_{z}, {S}^{+}, {S}^{-}\}$ and $\{{J}_{z}, {J}^{+}, {J}^{-}\}$ for spin and charge, respectively. These generators can be written in the following forms
\be\label{23}
{S}_{z}=\frac{1}{2}\sum_{ j\in \tilde{N }}({c}^{\dagger}_{j+}{c}_{j+}-{c}^{\dagger}_{j-}{c}_{j-}),\,\,{S}_{+}={{S}_{-}}^{\dagger}=\sum_{ j\in \tilde{N }}{c}^{\dagger}_{j+}{c}_{j-},
\ee
\be\label{24}
{J}_{z}=\frac{1}{2}\sum_{ j\in \tilde{N }}({c}^{\dagger}_{j+}{c}_{j+}+{c}^{\dagger}_{j-}{c}_{j-}-1),\,\,{J}_{+}={{J}_{-}}^{\dagger}=\sum_{ j\in \tilde{N }}(-1)^{j}{c}^{\dagger}_{j+}{c}^{\dagger}_{j-},
\ee
and the transfer between these two sets is known as the Shiba transformation \cite{Lieb, essler}.\\  
The total number of particles $N_{e}$, taken as the eigenvalue of the operator ${N}_{e}=\sum_{j\in \tilde{N}}({n}_{j+}+{n}_{j-})$, together with the number of particles $N_{i}$, with given one-node spin projection $i\in\{+,-\}$, taken as the eigenvalues of the operators ${N}_{i}=\sum_{j\in \tilde{N}}{n}_{ji}$, are quantum numbers. This mean the conservation of the total magnetization, given as the eigenvalue of the operator ${S_{z}}$. Furthermore as the result of the commutations of the Hamiltonian with the operators $(\ref{23})$ and $(\ref{24})$ (at half-filling) \cite{Cuoco} three additional quantum numbers appear, what together with $S_{z}$ gives the following set $\{S_{z}, J_{z}, S, J\}$, where $S$ and $J$ derive from the eigenvalues of the operators
\be\label{27}
({\bf S})^{2}=\frac{1}{2}[({S}_{+})^{2}+({S}_{-})^{2}]+({S}_{z})^{2},\,\,({\bf J})^{2}=\frac{1}{2}[({J}_{+})^{2}+({J}_{-})^{2}]+({J}_{z})^{2},
\ee
respectively, and create the appropriate basis called the \emph{spin basis}.



\section{The Schur-Weyl duality}

To be able to apply the SWD \cite{jak2018, Pjak} to the system of $N$ spins $s$ we need to define representation space for both symmetric $\Sigma_N$ and unitary $U(n)$ groups. Thus, we identify the single-node spin space with the space $h \cong \mathbb C_{n}, \quad n = 2s + $ 1 which is a natural space representation for the algebra $ \mathbb{C} [U(n)] $ of the unitary group $U(n)$.
Then, the Hilbert space of the system of a form $\mathcal{H} = h^{\otimes N}$ is the representation space of $\Sigma_N$ and $ U(n)$ groups.
Symmetric group acts naturally on the $N$-fold tensor product $h^{\otimes N}$, by permuting tensor factors, whereas unitary group performs unitary rotations of each factor. These two actions mutually commute, which generates a two-module structure of $\mathbb {C}[U(n)]\times\mathbb{C}[\Sigma_N] $ in Hilbert space $\mathcal{H}$ of the system. Thus, according to the SWD, if we define algebra representation as follows
\be
\mathbb{C}[U(n)] \stackrel{B}{\longrightarrow} \mbox{End}(\mathcal H) \stackrel{A}{\longleftarrow} \mathbb{C}[\Sigma_N]
\ee
then the following dependencies take place
\be
\begin{array}{l}
B(\mathbb{C}[U(n)])= \mbox{End}_{\Sigma_N}(\mathcal H),\\
\\
A(\mathbb{C}[\Sigma_N])= \mbox{End}_{U(n)}(\mathcal H),\\
\end{array}
\ee
where $\mbox{End}_{\Sigma_N}(\mathcal H)$ denotes the set of linear endomorphisms of $\mathcal{H}$ commuting with each endomorphism coming from $\Sigma_N$ group, and $\mbox{End}_{U(n)}(\mathcal{H})$ is a set of linear endomorphisms of $\mathcal{H}$, commuting with every endomorphism coming from $U(n)$. 
The above centralization of the algebras, allows to connect representation theory of the unitary group
with that of the symmetric group. The last one provides the unique and powerful combinatoric tools, which are very useful in the study of physical properties of one-dimensional spin systems.
The main outcome of the application of SWD, is a canonical decomposition of the Hilbert space of the spin system, into simple non-isomorphic modules
\be\label{rozklad_SW}
\mathcal{H}=\bigoplus_{\lambda \vdash N} (U^\lambda \otimes V^\lambda),
\ee
where $U^\lambda$ and $V^\lambda$ are simple modules of $U(n)$ and $\Sigma_N$, respectively. The sum runs over all partitions $\lambda$ of the integer $N$.
The above approach is a main idea which allows us to apply SWD to the particular physical systems.\\
The Hilbert space $\mathcal{H}$ is the scene of two dual actions, the symmetric group
$A: \Sigma_N \times \mathcal{H} \rightarrow \mathcal{H}$ and unitary group
$ B: U(n) \times \mathcal{H} \rightarrow \mathcal{H}$, determined on the computational basis vectors (\ref{13}).
The representation $A$ is given by the permutation of nodes
$
A(\sigma) |f \rangle = | i_{\sigma^{-1}(1)}, ..., i_{\sigma^{-1}(N)} \rangle, f \in  \tilde{n}^{\tilde N},  \sigma \in \Sigma_N,
$ while $B$ is given by simultaneous unitary rotations in single-node spaces
$
B(u) |f \rangle = | u i_1, ..., u i_N \rangle,  u \in U(n).
$
It is obvious, that those two actions mutually commute, i.e.
$
[A(\sigma),B(u)]=0,
$
for each $\sigma \in \Sigma_N$ i $u \in U(n)$.
Therefore, according to Heisenberg uncertainty principle, appropriate observables connected with those two actions can be simultanously "measureable".
Maximum set of such observables, can be implemented in the irreducible basis of the space  $\mathcal{H}$,  adjusted to the symmetry of the model.
To determine this set, the actions $A$ and $B$ should be decomposed into appropriate irreps
\be\label{r15}
A=\sum_{\lambda \in D_W(N,n)}\,\, m(A,\Delta^{\lambda})\,\,\Delta ^{\lambda},
\quad
B=\sum_{\lambda \in D_W(N,n)}\,\, m(B,D^\lambda)\,\,D^{\lambda},
\ee
where $\Delta^\lambda$ and $D^\lambda$ labeled by partition $\lambda$ denote irreps of the symmetric and unitary group, respectively. The sum runs over all partitions $\lambda$ belonging to the set $D_W(N,n)$ of all partitions of the number $N$ into no more than $n$ parts
and $m(A,\Delta^\lambda)$ and $m(B,D^\lambda)$ are the appropriate multiplicities.
Taking into account the SWD, one can write down the following relations
\be\label{r17}
m(A,\Delta^\lambda)=\mbox{dim} \, D^{\lambda},
\quad
m(B,D^\lambda)=\mbox{dim} \, \Delta^{\lambda}.
\ee
It means that the multiplicity of occurrence of irrep $\Delta^{\lambda}$ in $A$, is equal to the dimension of irrep $D^{\lambda}$, while the multiplicity of occurrence of irrep $D^\lambda$ in $B$ is equal to the dimension of irrep $\Delta^\lambda$.
The equations (\ref{r17}) results in decomposition of Hilbert space into the direct sum
\be\label{r19}
\mathcal{H} = \bigoplus_{\lambda \in D_W(N,n)}\mathcal{H} ^{\lambda}
\ee
of sectors $\mathcal{H} ^{\lambda}$, whereas each sector is divided into the direct product of modules 
\be\label{r22}
\mathcal{H}^\lambda = U^\lambda \otimes V^\lambda
\ee
of unitary and symmetric group, respectively.

\section{The Schur-Weyl duality for one-dimensional Hubbard model in the case of half-filling}

The action
\be\label{A}
A: \Sigma_N \times \tilde 4^{\tilde N} \longrightarrow \tilde 4^{\tilde N}
\ee
of the symmetric group $\Sigma_{N}$ on the set $\tilde{4}^{\tilde N}$ provides the orbits $\mathcal{O}_\mu $ of the group $\Sigma_{N}$ labeled by the \emph{weight} $\mu$, given as the sequence of non-negative integers 
\be\label{mu}
\mu = (\mu_1, \mu_2, \mu_3, \mu_4),
\ee
where the consecutive $\mu_{i}$ denote the number of $+$, $-$, $\pm$ and $\emptyset$ in the electron configuration, respectively, with relation $\sum_{i\in \tilde 4} \mu_i = N$, defined by the following equation
\be\label{31}
\mu_i = |\{ i_j = i \, | \, j \in \tilde N  \}|, \,\,\, i \in \tilde 4.
\ee
Such an orbit is invariant under the action of the symmetric group $\Sigma_{N}$ and forms the carrier space of the transitive representation $R^{\Sigma_{N}:\Sigma^{\mu}}$, with the stabilizer $\Sigma^{\mu}$ being the Young subgroup $\Sigma^{\mu}=\Sigma_{\mu_{1}}\times \Sigma_{\mu_{2}}\times \Sigma_{\mu_{3}}\times \Sigma_{\mu_{4}}$, {where} $\times$ denotes the Cartesian product.

Since there are two independent $SU(2)$ symmetries one can consider the action of the symmetric group $\Sigma_{N}$ - in context of the Schur-Weyl duality \cite{jak2018} - separately in the spin and pseudo-spin space in order to obtain the total spin $S$ and the total pseudo-spin $J$. This observation holds for the half-filled system of any even number $N$ of atoms and for open boundaries \cite{essler} for $N$ odd, and in both cases provides two symmetric groups $\Sigma_{N'}$ and $\Sigma_{N''}$ in the spin and pseudo-spin space, respectively. 
The actions 
\be
A: \Sigma_N \times \tilde 4^{\tilde N} \longrightarrow \tilde 4^{\tilde N},\,\,\,B: U(4) \times \tilde 4^{\tilde N} \longrightarrow \tilde 4^{\tilde N}
\ee
are replaced by
\be\label{a1}
A': \Sigma_{N'} \times \tilde 2^{\tilde N'} \longrightarrow \tilde 2^{\tilde N'}, \,\,\,B': SU(2) \times \tilde 2^{\tilde N'} \longrightarrow \tilde 2^{\tilde N'},
\ee
in the spin space $\mathcal{H}_{s}=lc_{\mathbb{C}} \,\, \tilde{2}^{\tilde {N}'}=h_{s}^{\otimes N'}$, where $h_{s}\cong \mathbb{C}_{2}$ denotes the one-node spin space, and
\be\label{a2}
A'': \Sigma_{N''} \times \tilde 2^{\tilde N''} \longrightarrow \tilde 2^{\tilde N''}, \,\,\,B'': SU(2) \times \tilde 2^{\tilde N''} \longrightarrow \tilde 2^{\tilde N''},
\ee
in the pseudo-spin space  $\mathcal{H}_{p}=lc_{\mathbb{C}} \,\, \tilde{2}^{\tilde {N}''}=h_{p}^{\otimes N''}$, where $h_{p}\cong \mathbb{C}_{2}$ denotes the one-node pseudo-spin space. The spin and pseudo-spin space are isomorphic with Hilbert space of the one-dimensional Heisenberg model for the case of $N'$ and $N''$ nodes of the spin chain, respectively, thus, the results presented in section {\bf 3} can be used not only for the single-node spin space but also for the single-node pseudo-spin space.

Let define some initial Hilbert space as follows
\be\label{total Hilbert}
\begin{array}{c}
\mathcal{H}_{int}=\bigoplus_{(\tilde{N'},\tilde{N''})} \left(\mathcal{H}_{s}\otimes \mathcal{H}_{p}\right) \cong \sum_{(\tilde{N}',\tilde{N}'')}\bigoplus \left[ (\mathbb{C}_{2})^{\otimes N'}\otimes (\mathbb{C}_{2})^{\otimes N''}\right],\\\\
\tilde{N}'\cup\tilde{N}''=\tilde{N}, \tilde{N}'\cap\tilde{N}''=\emptyset,\\
\end{array}
\ee
where $N'$ and $N''$ denotes the cardinalities of the sets $\tilde{N}'$ and $\tilde{N}''$, respectively, and $(\tilde{N}',\tilde{N}'')$ stands for the pair of these two sets - each taken in ascending order. The last equation means that from now on we will label the Hilbert space $(\ref{14})$ by $\mathcal{H}_{int}$. The space $(\ref{total Hilbert})$ can be decomposed with respect to the number of electrons in the system
\be
\mathcal{H}_{int}=\bigoplus_{N_{e}=0}^{2N}\mathcal{H}^{N_{e}},
\ee
and further with respect to the number of electrons with fixed spin projection
\be
\mathcal{H}^{N_{e}}=\bigoplus_{(N_{+},N_{-})}\mathcal{H}^{N_{e}}_{(N_{+},N_{-})},\,\,N_{+}+N_{-}=N_{e}.
\ee
Since the symmetry $SU(2)\times SU(2)$ holds only for the half-filling case the subspace $\mathcal{H}^{N_{e}=N}\equiv \mathcal{H}$ of the initial space $(\ref{total Hilbert})$ provides the proper Hilbert space $\mathcal{H}$ for the case considered in the present paper.\\ 
The actions $(\ref{a1})$ and $(\ref{a2})$ provide two transitive representations
\be
R^{\Sigma_{N'}:(\Sigma_{\mu_{1}}\times \Sigma_{\mu_{2}})}\,\,\mbox{and}\,\,R^{\Sigma_{N''}:(\Sigma_{\mu_{3}}\times \Sigma_{\mu_{4}})}
\ee
in the spin and pseudo-spin space, respectively, where $\Sigma^{\mu '}=\Sigma_{\mu_{1}}\times \Sigma_{\mu_{2}}$ and $\Sigma^{\mu ''}=\Sigma_{\mu_{3}}\times \Sigma_{\mu_{4}}$.\\ 
Each transitive representation in the spin space decomposes as follows
\be\label{trans_spin}
R^{\Sigma_{N'} : \Sigma^{\mu '}} \cong \sum_{ {\lambda '} \unrhd {\mu '}} K_{{\lambda '} \, {\mu '}} \,\, \Delta^{\lambda '}=\sum_{ {\lambda '} \unrhd {\mu '}} \Delta^{\lambda '}
\ee
into irreps of the symmetric group $\Sigma_{N'}$, with the partition ${\lambda '} \vdash {N'}$ defining the shape of the corresponding irrep $\Delta^{\lambda '}$, where $K_{\lambda ' \, \mu '}$ are the famous Kostka numbers, equal to 1 in case of two-dimensional one-node space, the sum runs over all partitions ${\lambda '}$ of $N'$ which are not smaller than $\mu '$ in the dominance order, and $N'$ denotes the number of appropriate one-node spin spaces $h_{s}$. The decomposition $(\ref{trans_spin})$ can be rewritten in more details as presented below
\be\label{trans_spin1}
R^{\Sigma_{N'} : (\Sigma_{N'-{\mu _{2}}}\times \Sigma_{\mu _{2}})}\cong \sum_{r=0}^{\mu_{2}}\Delta^{\{N'-r,r\}},
\ee
and the total spin $S$ and the magnetization $S_{z}$ have the following forms
\be\label{S}
S=\frac{N'}{2}-r,\,\,\,0\leq r \leq \mu_{2},
\ee
\be\label{Sz}
S_{z}=\frac{N'}{2}-\mu_{2}.
\ee
For the pseudo-spin space with $N''$ number of appropriate one-node pseudo-spin spaces $h_{p}$ by analogy to ($\ref{trans_spin}$) the following decomposition holds 
\be\label{trans_pseudo_spin}
R^{\Sigma_{N''} : \Sigma^{\mu ''}} \cong \sum_{ {\lambda ''} \unrhd {\mu ''}} K_{{\lambda ''} \, {\mu ''}} \,\, \Delta^{\lambda ''}=\sum_{ {\lambda ''} \unrhd {\mu ''}} \Delta^{\lambda ''}
\ee
into irreps of the symmetric group $\Sigma_{N''}$, with the partition ${\lambda ''} \vdash {N''}$ defining the shape of the corresponding irrep $\Delta^{\lambda ''}$. In analogy to $(\ref{trans_spin1})$ the decomposition $(\ref{trans_pseudo_spin})$ can be rewritten as follows
\be\label{trans_pseudo_spin1}
R^{\Sigma_{N''} : \left(\Sigma_{\frac{N''}{2}}\times \Sigma_{\frac{N''}{2}}\right)}\cong \sum_{r=0}^{\frac{N''}{2}}\Delta^{\{N''-r,r\}},
\ee
and the total pseudo-spin $J$ and $J_{z}$ have the following forms
\be\label{J}
J=\frac{N''}{2}-r,\,\,\,0\leq r \leq \frac{N''}{2},
\ee
\be\label{Jz}
J_{z}=\frac{N''}{2}-\mu_{4}=0.
\ee
For the half-filling case the number $N''$ is always even, $\mu_{3}=\mu_{4}=\frac{N''}{2}$, the quantum number $J_{z}$ is equal to $0$ and the eigenvalues of the operator $J_{z}$ are within the set $\{0,2,...,\frac{N''}{2}\}$.

\section{Jucys-Murphy operators}

In order to calculate the irreducible basis of the symmetric group for any number of atoms in the spin and pseudo-spin spaces for the half-filling of the electrons one can use the technique of Jucys-Murphy operators $\hat{M}_{j}$. These operators defined within the symmetric group algebra $\mathbb{C}[\Sigma_{N}]$ as the sum of all transpositions $(j,j')$ of the node $j\in \tilde{N}$ with preceding nodes $j'<j$, are introduced by Jucys \cite{1,2} and independently by Murphy \cite{3}, thus, they are called Jucys-Murphy operators. These $N-1$ hermitian and mutually commuting operators of the form
\be\label{9}
\hat{M}_{j}=\sum_{1\leq j' <j}(j,j'),\,\, j=2,3,\dots , N,
\ee
generate a maximal Abelian subalgebra in $\mathbb{C}[\Sigma_{N}]$. The standard Young tableaux $|\lambda \, y>$ \cite{sagan} of the shape $\lambda\vdash N$, i.e. the tableaux of this shape in the alphabet $\tilde N$ of atoms, with strictly increasing entries in rows and columns, constitutes the common eigenvector $|\lambda \, y>$ of the set of $\hat{M}_{j}$ operators, that is
\be\label{9'}
\hat{M}_{j}|\lambda \, y>=m_{j}(y)|\lambda \, y>,
\ee
with eigenvalues
\be\label{9''}
m_{j}(y)=c_{j}(y)-r_{j}(y),
\ee 
where the pair of positive integers $(c_{j}(y),r_{j}(y))$ gives the positions (the column and the row) of the number $j$ in tableaux $|\lambda \, y>$. In this way each basis function of the irreducible representation $\Delta^{\lambda}$ of the symmetric group $\Sigma_{N}$, labeled by the Young tableaux $|\lambda \, y>$, can be completely determined by the sequence $(m_{1}=1, m_{2},\dots , m_{N})$ of eigenvalues $(\ref{9''})$. The realization of each such irreducible vector within the group algebra $\mathbb{C}[\Sigma_{N}]$ is given via the projector operator $e_{yy}^{\lambda}$ of the well known Young orthogonal basis \cite{1,2,3}, where the remarkable significance of the Jucys-Murphy operators is underlined
\be
e_{yy}^{\lambda} = | \lambda w y\rangle \langle \lambda w y|, 
\ee
with $w$ denoting appropriate repetition label. 
Thus
\be\label{p}
e_{yy}^{\lambda}=\prod_{j=2}^{N}\prod_{\{y_{j-1}|y_{j-1}^{+}\neq y_{j}\}}\frac{\hat{M}_{j}-m_{j}(y_{j-1}^{+})\hat{I}}{m_{j}(y)-m_{j}(y_{j-1}^{+})},
\ee
where $y \in SYT(\lambda)$, with $SYT(\lambda)$ being the set of all standard Young tableaux of the shape $\lambda$, $y_{j}$ denotes the tableaux obtaining from $y$ by extracting the set $\{j+1, j+2, \dots N\}$ of numbers, $y_{j-1}^{+}$ can be created from $y_{j-1}$ after adding to its entries the number $j$, and $\hat{I}$ stands for the appropriate unity operator.

\section{The example of the chain consisted of eight atoms.}

Now we want to discus the case of the chain consisted of eight atoms in the restriction of the half-filling of electrons. Since the numbers of up- and down-spin electrons are separately conserved the Hamiltonian $(\ref{11})$ gets reduced to diagonalizing it in sectors characterized by elements of the subset
\be\label{77}
\{(8,0),(7,1),(6,2),(5,3),(4,4),(3,5),(2,6),(1,7),(0,8)\}
\ee
of the cartesian  product $N_{+} \times N_{-}$. 
The dimension of the initial space $(\ref{total Hilbert})$ is equal to
\be
\begin{array}{c}
\mbox{dim}\,\mathcal{H}_{int}=\mbox{dim}\,\bigoplus_{N_{e}=0}^{16}\mathcal{H}^{N_{e}}=\\\\
=1+16+120+560+1820+4368+8008+11440+\\\\
+12870+11440+8008+4368+1820+560+120+16+1=\\\\
=65\,536=4^{8}.\\
\end{array}
\ee
The dimension of the proper Hilbert space given as the subspace $\mathcal{H}^{N_{e}=N=8}$ of the initial space $(\ref{total Hilbert})$
\be
\mathcal{H}=\mathcal{H}^{N_{e}=8}=\bigoplus_{(N_{+},N_{-})}\mathcal{H}^{N_{e}=8}_{(N_{+},N_{-})},\,\,N_{+}+N_{-}=8,
\ee
where the sum runs through the elements of the set $(\ref{77})$, can be calculated as follows
\be
\begin{array}{c}
\mbox{dim}\, \mathcal{H}=\mbox{dim}\, \mathcal{H}^{8}_{(8,0)}+\mbox{dim}\, \mathcal{H}^{8}_{(7,1)}+\mbox{dim}\, \mathcal{H}^{8}_{(6,2)}+\mbox{dim} \,\mathcal{H}^{8}_{(5,3)}+\mbox{dim} \,\mathcal{H}^{8}_{(4,4)}+\\\\
\mbox{dim} \,\mathcal{H}^{8}_{(3,5)}+\mbox{dim} \,\mathcal{H}^{8}_{(2,6)}+\mbox{dim}\,\mathcal{H}^{8}_{(1,7)}+\mbox{dim} \,\mathcal{H}^{8}_{(0,8)},\\\\
\mbox{dim}\, \mathcal{H}=1+64+784+3136+4900+3136+784+64+1=12\,870.\\
\end{array}
\ee
Using SWD one can assign the value of the total spin $(\ref{S})$ to each irrep of the decomposition $(\ref{trans_spin1})$ and the value of the total pseudo-spin $(\ref{J})$ to each irrep of the decomposition $(\ref{trans_pseudo_spin1})$, what presents Table $\ref{Tab.1}$.
\begin{table}
\begin{tabular}{|c|c|c|c|c|c|c|c|c|c|c|}
\hline
$\mu' $&$\mu''$&$\lambda'$&$\lambda''$&$S$&$J$&dim$ \Delta^{\lambda '}$&dim$ \Delta^{\lambda ''}$&$\tau$&$x_{\mu}$\\
\hline
$(5,3)$&$(0,0)$&$\{8\}$&$-$&$4$&$-$&1&$-$&1&56\\
&&$\{7\,1\}$&$-$&$3$&$-$&7&$-$&&\\
&&$\{6\,2\}$&$-$&$2$&$-$&20&$-$&&\\
&&$\{5\,3\}$&$-$&$1$&$-$&28&$-$&&\\
\hline
$(4,2)$&$(1,1)$&$\{6\}$&$\{2\}$&$3$&1&1&1&28&840\\
&&$\{5\,1\}$&$\{1^2\}$&$2$&0&5&1&&\\
&&$\{4\,2\}$&&$1$&&9&&&\\
\hline
$(3,1)$&$(2,2)$&$\{4\}$&$\{4\}$&$2$&2&1&1&70&1\,680\\
&&$\{3\,1\}$&$\{3\,1\}$&$1$&1&3&3&&\\
&&&$\{2^2\}$&&0&&2&&\\
\hline
$(2,0)$&$(3,3)$&$\{2\}$&$\{6\}$&$1$&3&1&1&28&560\\
&&&$\{5\,1\}$&&2&&5&&\\
&&&$\{4\,2\}$&&1&&9&&\\
&&&$\{3^2\}$&&0&&5&&\\
\hline
\end{tabular}
\caption{The total number of $3\,136$ states for the case of $N_{+}=5$, $N_{-}=3$, $S_{z}=1$, $J_{z}=0$, and $x_{\mu}=\tau \cdot \sum_{{\lambda '} \unrhd {\mu '}}$dim$ \Delta^{\lambda '}\cdot \sum_{{\lambda ''} \unrhd {\mu ''}}$dim$ \Delta^{\lambda ''}$.}\label{Tab.1}
\end{table}
The number $\tau$ denotes the multiplicity of deploying of $N'$ one-node spin spaces $h_{s}$ and $N''$ one-node pseudo-spin spaces $h_{p}$ on  $N=N'+N''$ positions and is equal to
\be\label{tau}
\tau =\left (
\begin{array}{c}
N\\
N'\\
\end{array}
\right )=\left (
\begin{array}{c}
N\\
N''\\
\end{array}
\right ).
\ee
Table $\ref{Tab.2}$ presents the example of using the operators $(\ref{p})$ in order to obtain the irreducible basis of the symmetric group $\Sigma_{4}$ taken in the basis of the electron configurations for the case of $\mu_{1}=\mu_{2}=2$ and for the orbits $\mathcal{O}_{++--}$ and $\mathcal{O}_{+-+-}$ of the cyclic group $C_{4}$.
\begin{table}
\begin{tabular}{c|c|c|c|c|c|c}
$f \setminus  y$ &\young(1234)&\young(123,4)&\young(134,2)&\young(124,3)&\young(12,34)&\young(13,24)\\
\hline
$|++--\rangle $ & $\frac{\sqrt{6}}{6} $ & $-\frac{\sqrt{6}}{6} $ & $0 $ & $-\frac{\sqrt{3}}{3} $ & $\frac{\sqrt{3}}{3} $ & $0$\\
$|-++-\rangle $ & $\frac{\sqrt{6}}{6} $ & $-\frac{\sqrt{6}}{6} $ & $\frac{1}{2} $ & $\frac{\sqrt{3}}{6} $ & $-\frac{\sqrt{3}}{6} $ & $-\frac{{1}}{2}$\\
$|--++\rangle $ & $\frac{\sqrt{6}}{6} $ & $\frac{\sqrt{6}}{6} $ & $0$ & $\frac{\sqrt{3}}{3} $ & $\frac{\sqrt{3}}{3} $ & $0$\\
$|+--+\rangle $ & $\frac{\sqrt{6}}{6} $ & $\frac{\sqrt{6}}{6} $ & $-\frac{1}{2} $ & $-\frac{\sqrt{3}}{6} $ & $-\frac{\sqrt{3}}{6} $ & $-\frac{{1}}{2}$\\
$|+-+-\rangle $ & $\frac{\sqrt{6}}{6} $ & $-\frac{\sqrt{6}}{6} $ & $-\frac{1}{2} $ & $\frac{\sqrt{3}}{6} $ & $-\frac{\sqrt{3}}{6} $ & $\frac{{1}}{2}$\\
$|-+-+\rangle $ & $\frac{\sqrt{6}}{6} $ & $\frac{\sqrt{6}}{6} $ & $\frac{1}{2} $ & $-\frac{\sqrt{3}}{6} $ & $-\frac{\sqrt{3}}{6} $ & $\frac{{1}}{2}$\\
\end{tabular}
\caption{The irreducible basis of the symmetric group $\Sigma_{4}$ taken in the basis of the electron configurations for the case of $N_{+}=N_{-}=2$ and for the orbits $\mathcal{O}_{++--}$ and $\mathcal{O}_{+-+-}$ of the cyclic group $C_{4}$.}\label{Tab.2}
\end{table}

\section{Conclusions}

We presented the application of the Schur-Weyl duality in the one-dimensional Hubbard model in the case of half-filling for any number of atoms. We showed the way of using the Schur-Weyl duality in spin and pseudo-spin space in order to obtain the total spin $S$ and the total pseudo-spin $J$ (Table $\ref{Tab.1}$). We created both the spin and pseudo-spin spaces as the appropriate tensor product of the one-node spin and pseudo-spin spaces and provide the space of all quantum states of the considered system taken over all possible locations of these spaces on the $N$- atoms chain. We used the concept of the initial Hilbert space which provides the proper Hilbert space of the considered system as its subspace, since there is the confinement of half-filling. We gave the expressions for the total spin $S$ and the total pseudo-spin $J$ in context of the representation theory of the symmetric group.

The calculations are significant since there is a lack of analytical calculations in the literature of using of the symmetry $SU(2)\times SU(2)$, which is crucial in understanding the Hubbard model. The obtained results lead to a significant reduction in the size of the Hubbard Hamiltonian and can easily be implemented into numerical simulations due to the use of simple transpositions being the generators of the symmetric group. 

The discussed issues are important since they concern finding the accurate solutions of spin and electron models with effective and simple methods, and they are necessary when it comes to the development of quantum algorithms based on binary functions, whose arguments are often only a few bits. In order to further develop the science in the field of quantum computers and create new algorithms for solving more and more difficult tasks, one should understand the quantum mechanics of spin and electron systems composed of several particles in order to later generalize the considerations into larger dimensions and a larger number of particles.


\providecommand{\href}[2]{#2}

\address{
Faculty of Mathematics and Applied Physics,\\
Rzesz\'{o}w University of Technology,\\
al. Powsta\'nc\'ow Warszawy 12, 35-959 Rzesz\'ow,Poland\\
\email{djak@prz.edu.pl}\\
}


\begin{thebibliography}{10}

\bibitem{Hubbard}
J. Hubbard, {\em {Electron correlations in narrow energy bands}}. {Proc. Roy. Soc. London Ser. A}, {\bf 276}, 238--257, (1963), \href{ https://doi.org/10.1098/rspa.1963.0204}{{https://doi.org/10.1098/rspa.1963.0204}}. 

\bibitem{essler}
F. H. L. Essler et al., {\em {The One-Dimensional Hubbard Model}}. {Cambridge University Press}, (2005), \href{https://doi.org/10.1017/CBO9780511534843}{{https://doi.org/10.1017/CBO9780511534843}}. 

\bibitem{Gebhard}
F. Gebhard, {\em {The Mott Metal-Insulator Transition}}. {Springer}, (1997), \href{https://doi.org/10.1007/3-540-14858-2}{{https://doi.org/10.1007/3-540-14858-2}}. 

\bibitem{jak2018}
D. Jakubczyk, P. Jakubczyk, {\em {Combinatorial approach to the representation of the Schur-Weyl duality in one-dimensional spin systems
}}. {J. Math. Phys.}, {\bf 59}, 023504--10, (2018), \href{ https://doi.org/10.1063/1.5006328
}{{ https://doi.org/10.1063/1.5006328}}.

\bibitem{qudit}
P. Jakubczyk, S. Topolewicz, A. Wal and T. Lulek, {\em {Magnonic qudit and algebraic Bethe Ansatz
}}. {OSID}, {\bf 16}, 221--227 (2009), \href{ https://doi.org/10.1088/1742-6596/213/1/012009
}{{https://doi.org/10.1088/1742-6596/213/1/012009}}.

\bibitem{Schur}
I. Schur, {\em {Uber die rationalen Darstellungen der allgemeinen linearen Gruppe}}. {Sitzungsberichte Akad. Berlin}, 58--75, (1927).

\bibitem{Weyl}
H. Weyl, {\em {The Classical Groups. Their Invariants and Representations}}. {Princeton UP, Princeton, N.J}, (1946).

\bibitem{per}
D. Jakubczyk, P. Jakubczyk, {\em {On the permutational symmetry of the Hubbard model}}. {Acta Phyica Polonica B},  {\bf 42}, 1825--1836, (2011), \href{https://doi.org/10.5506/APhysPolB.42.1825}{{https://doi.org/10.5506/APhysPolB.42.1825}}.
 
\bibitem{jakubczyk1}
D. Jakubczyk, P. Jakubczyk, {\em {On the $SU(2)\times SU(2)$ symmetry in the Hubbard model}}. {Cent. Eur. J. Phys.}, {\bf 10}, 906--912, (2012), \href{https://doi.org/10.2478/s11534-012-0055-6}{{https://doi.org/10.2478/s11534-012-0055-6}}.

\bibitem{Lieb}
E. H. Lieb, F. Y. Wu, {\em {Absence of Mott Transition in an Exact Solution of the Short-Range, One-Band Model in One Dimension}}. {Phys. Rev. Let.}, {\bf 20}, 1445--1448 (1968), \href{https://doi.org/10.1103/PhysRevLett.20.1445
}{{https://doi.org/10.1103/PhysRevLett.20.1445}}.

\bibitem{Yang1}
C. N. Yang, C. P. Yang, {\em {One-Dimensional Chain of Anisotropic Spin-Spin Interactions. II. Properties of the Ground-State Energy Per Lattice Site for an Infinite System}}. {Phys. Rev. Lett.}, {\bf 150}, 327--339, (1966), \href{https://doi.org/10.1103/PhysRev.150.327
}{{https://doi.org/10.1103/PhysRev.150.327}}.

\bibitem{Yang2}
C. N. Yang, {\em {Some Exact Results for the Many-Body Problem in one Dimension with Repulsive Delta-Function Interaction}}. {Phys. Rev. Lett.}, {\bf 19}, 1312--1315, (1967), \href{https://doi.org/10.1103/PhysRevLett.19.1312
}{{https://doi.org/10.1103/PhysRevLett.19.1312}}.

\bibitem{Yang3}
C. N. Yang, {\em {$\eta $ pairing and off-diagonal long-range order in a Hubbard model}}. {Phys. Rev. Lett.}, {\bf 63}, 2144--2147, (1989), \href{https://doi.org/10.1103/PhysRevLett.63.2144}{{https://doi.org/10.1103/PhysRevLett.63.2144}}.

\bibitem{Yang4}
C. N. Yang, {\em {$SO_{4}$ symmetry in a Hubbard model}}. {Mod. Phys. Lett. B}, {\bf 4}, 759--766, (1990), \href{https://doi.org/10.1142/S0217984990000933}{{https://doi.org/10.1142/S0217984990000933}}.

\bibitem{homotopia}
B. Lulek, D. Jakubczyk, {\em {Homotopy of classical configuration space for the two-magnon sector of a magnetic Heisenberg ring}}. {Cent. Eur. J. Phys.}, {\bf 1}, 132--144, (2003), \href{https://doi.org/10.2478/BF02475557}{{https://doi.org/10.2478/BF02475557}}.

\bibitem{bazaprofili}
B. Lulek, T. Lulek, A. Wal, P. Jakubczyk, {\em {The basis of wavelets for a finite Heisenberg magnet}}. {Phys. B}, {\bf 337}, 375--387, (2003), \href{https://doi.org/10.1016/S0921-4526(03)00430-7}{{https://doi.org/10.1016/S0921-4526(03)00430-7}}.

\bibitem{Cuoco}
M. Cuoco et al., {\em {Application of the Global $SO(4)$ Symmetry in the Diagonalization of Translationally Invariant Correlated Electron Models}}. {Int. J. Mod. Phys. B}, {\bf 11}, 2511--2532, (1997), \href{https://doi.org/10.1142/S0217979297001271}{{https://doi.org/10.1142/S0217979297001271}}.

\bibitem{Pjak}
P. Jakubczyk ,T. Lulek, D. Jakubczyk, B. Lulek, {\em {The duality of Weyl and linear extension of Kostka matrices
}}. {J. Phys. Conference series}, {\bf 30}, 203--208, (2006), \href{https://doi.org/10.1088/1742-6596/30/1/024}{{https://doi.org/10.1088/1742-6596/30/1/024}}. 

\bibitem{1}
A. A. Jucys, {\em {On the Young operators of symmetric groups}}. {Litousk. Fiz Sb.}, {\bf 6}, 163--180, (1966).

\bibitem{2}
A. A. Jucys, {\em {Factorisation of Young's projection operators of symmetric groups}}. {Litousk. Fiz. Sb.}, {\bf 11}, 1--10, (1971).

\bibitem{3}
G. E. Murphy, {\em {The Idempotents of the Symmetric Groups and Nakayama's Conjecture}}. {J. Algebra}, {\bf 81}, 258--265, (1983), \href{https://doi.org/10.1016/0021-8693(83)90219-3}{{https://doi.org/10.1016/0021-8693(83)90219-3}}.

\bibitem{sagan}
 B. Sagan, {\em {The Symmetric Group: Representations, Combinatorial Algorithms, and Symmetric Functions}}. {Springer-Verlag}, New York, (2001).

\bibitem{bethe}
H. Bethe, {Z. Physik}, {\bf 71}, 205--226, (1931)(in German; English translation in: D.C. Mattis, {\em {The Many-Body Problem}}. {World Sci.}, Singapore, 689--716, (1993)), \href{http://dx.doi.org/10.1007/BF01341708}{{http://dx.doi.org/10.1007/BF01341708}}.




\end{thebibliography}
\end{document}